\documentclass{article}

\usepackage{amsmath,amssymb,upgreek,epsfig}

\newcommand{\x}{\mathbf{x}}
\renewcommand{\u}{\mathbf{u}}
\newcommand{\n}{\mathbf{n}}
\renewcommand{\v}{\mathbf{v}}
\newcommand{\f}{\mathbf{f}}
\newcommand{\dd}{\mathrm{d}}
\newcommand{\e}{\mathrm{e}}
\newcommand{\ii}{\mathrm{i}}
\newcommand{\D}{\mathrm{D}}

\newcommand{\C}{\mathsf{C}}
\renewcommand{\H}{\mathsf{H}}
\newcommand{\U}{\mathcal{U}}
\newcommand{\G}{\mathsf{G}}
\newcommand{\J}{\mathsf{J}}

\author{A. Hanyga}

\title{Elimination of memory from the 
equations of motion of hereditary viscoelasticity for increased efficiency
of numerical integration}

\begin{document}

\maketitle

\begin{abstract}
A method of eliminating the memory from the equations of motion of linear viscoelasticity
is presented. Replacing the unbounded memory by a quadrature over a finite or semi-finite interval leads to considerable reduction of computational effort and storage. The method applies
to viscoelastic media with separable completely monotonic relaxation moduli with 
an explicitly known retardation spectrum. In the 
seismological Strick-Mainardi model the quadrature is a Gauss-Jacobi quadrature on the interval$[-1,1]$. The relation to fractional-order viscoelasticity is shown.
\end{abstract} 

\section*{Notations.}

\begin{small}
\begin{tabular}{lll}
$\u\cdot \v$ & scalar product & $\sum_{l=1}^d u_l \, v_l$\\
$\langle \v, \mathsf{A} \u \rangle$ & & $\sum_{k,l}^d A_{klmn} \, v_{kl}\, u_{mn}$\\
$]a,b]$ & & $\{ x \mid a < x \leq b\}$ \\
$\tilde{f}(p)$ & Laplace transform & $\int_0^\infty f(t) \, \exp(-p t) \, \dd t$\\
$f\ast g(t)$ & Volterra convolution & $\int_0^t f(s)\, g(t-s)\, \dd s$\\
$\theta(x)$ & unit step function\\
\end{tabular}
\end{small}

\section{Introduction.}

Hereditary effects play an important role in continuum mechanics 
(viscoelasticity, poroelasticity) and in several other fields. Memory effects are 
implicitly present if the equations involve fractional derivatives, for example in
anomalous diffusion \cite{HanPRSL1}. The main disadvantage of hereditary models is the computational cost 
and storage problems. These problems are significant if fractional derivatives are approximated
according to the Gr\"{u}nwald-Letnikov formula. In the case of
fractional-order derivatives these difficulties have however been
circumvented by resorting to some integral representations \cite{Montseny,YuanAgrawal,
HanygaLu,LuHanyga,LuHanyga1,LuHanyga2,Diethelm2008,Diethelm2010}. It is our aim to demonstrate that
a similar integral representation is available for viscoelastic media with 
a completely monotonic relaxation modulus (or, equivalently, Bernstein creep compliance)
\cite{Molinari,HanDuality}. The main computational cost is then associated with a quadrature.
If the integration extends over an infinite interval then the accuracy vs cost ratio can be reduced by an explicit asymptotic estimate of the tail, as shown in the papers of Hanyga and Lu.
Alternatively, an infinite spectrum can be mapped onto a finite interval. A rigorous
analysis of this method and improvements are presented in \cite{Diethelm2010}.

For some viscoelastic equations of
physical interest the numerical scheme based on an integral representation is much more
accurate and efficient than in the case of fractional derivatives. This is true
for viscoelastic models with a finite retardation spectrum.  We shall consider a case
in which the quadrature can be reduced to a Gauss-Jacobi quadrature by a simple 
transformation of the integration variable. This model was proposed for seismological 
applications in \cite{StrickMainardi}. 

\section{Problem formulation.}

The Cauchy stress tensor $\upsigma$ is given by the constitutive equation
\begin{equation} \label{eq:const}
\upsigma(t,\x) = \int_0^t \G(s,\x)\, [\nabla \u^\prime(t - s,\x)]\, \dd s
\end{equation}
where the prime denotes the time derivative.
The relaxation modulus $\G$ is a function on $\mathbb{R}_+$ with values in the space $V$ of 
symmetric rank-2 operators on the space $S$ of symmetric tensors over $\mathbb{R}^d$. The action of an element $\mathsf{A}$  of $V$ on an element $\v$ of $S$ is denoted
by $\mathsf{A}[\v]$. 
We can extend an operator $\mathsf{A}$ to the space of all the rank-2 tensors over $\mathbb{R}^d$
by assuming that it vanishes on antisymmetric tensors. Applying this extension to 
the relaxation modulus, equation~\eqref{eq:const} can  
be expressed in the indicial notation
$$\sigma_{ij} = \int_0^t G_{ijkl}(s,\x) u^\prime_{k,l}(t - s,\x) \, \dd s$$
where $G_{ijkl}(s,\x) = G_{ijlk}(s,\x)$.  

\begin{gather} \label{eq:diff}
\rho \, \u^{\prime\prime}(t,\x)  = \nabla\cdot \left[ \G(t,\x)\ast\, \nabla \u^\prime(t,\x)\right] + \f(t,\x) \qquad t \geq 0, \x \in \U \\
\u(0,\x) = 0 , \quad \u^\prime(0,\x) = 0 \qquad \x \in \U \label{eq:IC}\\
\u(t,\x) \qquad\text{on $\mathbb{R}_+\times\Gamma_1$} \label{eq:BC1}\\
\n\cdot \G(0,\x)\,\nabla \u(t,\x) = 0 \qquad\text{on $\mathbb{R}_+\times\Gamma_2$} \label{eq:BC2}
\end{gather}
where $\Gamma_1 \cap \Gamma_2 = \empty$ and $\Gamma_1 \cup \Gamma_2 = \partial \U$. 
$\U$ is a open connected subset of $\mathbb{R}^d$ with a piecewise differentiable boundary 
$\partial \U$ and $\n$ is a unit outer normal on $\partial \U$. 
 
It is assumed that the function $\G$ is completely monotonic in the sense defined in 
\cite{HanDuality}, i. e. for every $v \in S$ the function 
$t \rightarrow \langle \v, \G(t,\x)\, \v\rangle \equiv v_{kl}\, G_{klmn}(t,\x)\, v_{mn}$ is 
completely monotonic \cite{BernsteinFunctions}. It is furthermore assumed that 
the function $\G$ is locally integrable. Every completely monotonic function is smooth
on $\mathbb{R}_+$, hence the only singularity is possible at 0. The function $\G$ 
is also positive semi-definite, i. e.  $\langle \v, \G(t)\, \v\rangle \geq 0$ for every 
$\v \in S$, hence the last assumption is equivalent to 
$\int_0^1 \G(t,\x) \, \dd t < \infty$. The locally integrable completely monotonic functions
will be referred to by the acronym LICM.

In \cite{HanDuality} it is proved that under our assumptions there is a unique function 
$\J: \mathbb{R}_+ \times \U \rightarrow V$, called the creep compliance, satisfying the equation
\begin{equation}\label{eq:duality}
\int_0^t \G(s,x) \, \J(t - s,x)\, \dd s = t\qquad t > 0
\end{equation}
(the function $\G$ can be replaced by $\J$ and vice versa). The integrand is a product of two operators for each $t > 0$ and $0 \leq s \leq t$.  In indicial notation 
$\int_0^t G_{ijkl}(s,\x) \,J_{klmn}(t - s,\x)\, \dd s = t\qquad t > 0$.
The function $\J(\cdot,\x)$ is a Bernstein function, i. e. it has a derivative which is a LICM 
function. The converse statement is also true, i. e. for every Bernstein function $\J$
there is a unique LICM function $\G$ satisfying equation~\eqref{eq:duality}. 
The function $\J$ can be extended to a continuous function $\J$ and
$0 \leq \J(0) < \infty$. The inequality $\mathsf{A} \leq \mathsf{B}$ for 
$\mathsf{A}, \mathsf{B} \in V$  is defined here 
by the relation $\langle \v, \mathsf{A} \, \v\rangle \leq \langle \v, \mathsf{B}\, \v \rangle $
for every $\v \in S$. It is also proved in \cite{HanDuality} that $\J(0) = 0$
if and only if the limit $\lim_{t\rightarrow 0} \G(t)$ is not finite.

The left-hand side is a Volterra convolution, which we shall denote as $\G \ast \J$.

\section{Case 1. General CM relaxation modulus, explicitly known relaxation spectrum.}

A LICM rank-4 tensor-valued relaxation modulus $\G$ has  
the integral representation
\begin{equation} \label{eq:rMCM}
\G(t) = \int_{[0,\infty[} \e^{-r t}\, \H(r) \, \mu(\dd r)
\end{equation}
where $\mu$ is a positive Radon measure on $[0,\infty[$ satisfying the inequality
\begin{equation}
\int_{[0,\infty[} \frac{\mu(\dd r)}{1 + r} < \infty
\end{equation}
while $\H$ is a rank-4 tensor-valued function on the support of $\mu$, which has the symmetries
of $\G$: $H_{klmn} = H_{klnm} = H_{mnkl}$, and is a positive semi-definite operator on $S$. The function $\H$ is bounded except for a set of 
$\mu$ measure 0 \cite{HanDuality}. 
In an inhomogeneous medium $\G$, $\H$ and $\mu$ can depend on $\x$, but we shall assume
here for simplicity that this is not the case.

Define the auxiliary variables
$$\Phi_{r;lmn}(t,\x) := \int_0^t \e^{-r (t-s)}\, u^{\;\prime}_{m,ln}(s,\x) \, \dd s $$
Substituting \eqref{eq:rMCM} in \eqref{eq:diff} yields the following system of
equations
\begin{gather} \label{eq:dd1}
\rho\, u_k^{\prime\prime} =  \int_{[0,\infty[} H_{klmn}(r) \, \Phi_{r;lmn}(t,\x),\qquad k=1,\ldots,d; t \geq 0; \x \in \U\\
\mathbf{\Phi}_r^{\;\prime}(t,\x) + r \, \mathbf{\Phi}_r(t,\x) = \nabla \nabla \u(t,\x)\qquad
r \geq 0; t \geq 0; \x \in \U \label{eq:dd2}
\end{gather}

Let $h$ be the step in time and $\u^i(\x) := \u(i h,\x)$, $\mathbf{\Phi}^i_r(\x) :=
\mathbf{\Phi}_r(i h,\x)$, $r \geq 0$, $\f^i(\x) := \f(i h,\x)$, $\x \in \U$. Assume a quadrature approximation 
\begin{equation}
\int_{[0,\infty[} f(r) \, \mu(\dd r) \sim \sum_{j=0}^N w_j\, f(r_j)
\end{equation}
with $w_j, r_j \geq 0$, $j = 1, \ldots, N$. 
Let $\mathbf{\Psi}_j(t,\x) := \mathbf{\Phi}_{r_j}(t,\x)$
Equations~(\ref{eq:dd1}--\ref{eq:dd2}) can be approximated by 
the following discretized system 
\begin{gather}
\rho\, \left[ u^{i+1}_k + u^{i-1}_k - 2 u^{i}_k\right] = h^2\, \sum_{j=1}^N w_j\,
H_{klmn}(r_j)\, \Psi^i_{j;lmn}(\x) + h \, f_k^i(\x)\qquad k = 1,\ldots, d;\;\x \in \U\\
\Psi^{i=1}_{j;lmn} - \Psi^i_{j;lmn} + h\, r_j \, \Psi^i{j;lmn} = h \, u^i_{m,ln} \qquad
j = 1,\ldots N; l,m,n = 1,\ldots, d; \x \in \U
\end{gather}
for i=0,1,2,\ldots.

\begin{figure}
\begin{center}
\epsfig{file=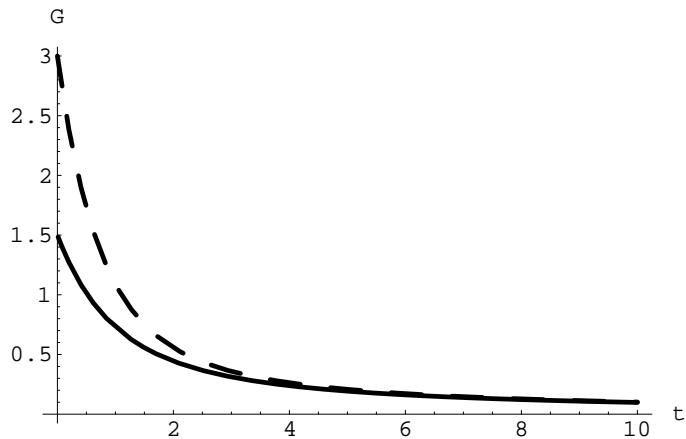,width=0.75\textwidth}
\end{center}
\caption{The function $g(t)$ for $\alpha = 1/3$ (solid curve) and $\alpha = -1/3$ (dashed curve).}
\label{fg}
\end{figure}

The above discretization is handy if the relaxation measure $\mu$ and the function $\H$ are
known The quadrature depends only on the measure $\mu$ because the function $\H$ is bounded on the support
of $\mu$.
The quadrature can often be evaluated with a few nodes instead of a long and growing number of memory points if a Gr\"{u}nwald-Letnikov discretization is applied. If the relaxation spectrum is unbounded then asymptotic estimates of the integrand can be helpful. In some cases the 
relaxation spectrum can be finite, for example if
$\mu(\dd r) = \theta(\Omega - r)\, (1 - r/\Omega)^\alpha$, $\alpha > -1$, $\Omega > 0$. Assuming 
that $\H(r) = \mathbf{Id}_S$, the relaxation modulus is a LICM function
$\G(t) = g(t) \, \mathrm{Id}_S$ with 
$g(t) = \Omega\, \e^{-\Omega\, t} (-\Omega t)^{-1-\alpha}\, [\Gamma(\alpha) - \Gamma(\alpha,-\Omega t)]$
(the two factors are complex valued, but the product is real valued). For $\alpha = 0.5$
$g(t) = 2/3 \, \e^{-\Omega t}\, _1F_1(1.5,2.5,\Omega t)$ and for $\alpha = -0.5$ 
$g(t) = 2 \, \e^{-\Omega t}\, _1F_1(0.5,1.5,\Omega t)$. For $\alpha = \pm 1/3$ the functions are shown in Fig.~\ref{fg}.
The integral over the relaxation spectrum $[0,\Omega]$ can be expressed in terms of the Gauss-Jacobi 
quadrature on $[-1,1]$ \cite{StroudSecrest}  by the formula
$$ \int_0^\Omega (1 - \Omega/r)^\alpha\, f(r)\, \dd r = \Omega\, 2^{-1-\alpha} 
\int_{-1}^1 (1 - x)^\alpha\, f(\Omega\, (x + 1)/2)\, \dd x$$
The quadrature nodes and weights can be calculated by the method presented in \emph{op. cit.} 

The relaxation spectral measures of several viscoelastic models of practical interest 
are presented in \cite{HanJMP2010,HanWM2013}. These include the Cole-Cole, Havriliak-Negami, Cole-Davidson models and Prony sums.  

\section{Case 2. Separable relaxation modulus, explicitly known retardation spectrum.} 
\label{retbased}

We shall show how the convolution operator can be eliminated/replaced by a quadrature.
The method is applicable to viscoelastic media with a separable
relaxation modulus 
\begin{equation} \label{eq:separability}
\G(s,\x) = g(s) \, \C(\x)
\end{equation}
where $\C(\x)$ is positive semi-definite symmetric operator on $S$, invertible for every 
$\x \in \U$ 
and $g$ is a LICM function. In this case equation~\eqref{eq:duality} is satisfied 
by $\J(s,\x) = j(s)\, \C(\x)^{-1}$, where $j(t)$ is the unique solution of the
equation
\begin{equation} \label{eq:dual1}
\int_0^t j(s) \, g(t - s) \, \dd s = t \qquad t > 0
\end{equation}
The function $j$ is non-decreasing and non-negative, hence it has a finite limit $j(0+) \geq 0$
as $t \rightarrow 0$. This limit is non-zero if $\lim_{t\rightarrow 0} g(t) < \infty$.
In this case $j(0+) = 1/g(0+)$ \cite{HanDuality}.

Equation~\eqref{eq:separability} means that the memory effects are independent of the 
material point and of the direction in an anisotropic medium. The anisotropy is entirely 
accounted for by the tensor $\C(\x)$. This condition is satisfied in particular by the 
shear flow of an 
isotropic viscoelastic medium. In this case $d = 1$, $\x$ reduces to a single coordinate
$x$, while $\u$ reduces to a component $u_y$ of $\u$ orthogonal to the $x$ direction. 
The tensor $\C$ can be set equal to 1 and $g(t)$ plays the role of the viscoelastic shear 
relaxation modulus, with $g(0) = \mu$ -- the Lam\'{e} elastic shear modulus. 

Assuming separability of the relaxation modulus, equation~\eqref{eq:diff} reduces to 
\begin{equation} \label{eq:diff-sep}
\rho\, \u^{\prime\prime}(t,\x) = g \ast  L\left[\u^\prime\right] + 
\f(t,\x) 
\end{equation}
where $L[\u] := \nabla\cdot \{\C(\x)[\nabla\, \u(t,\x)]\}$.

We shall now eliminate the convolution operator $g \ast$ by applying the convolution operator
$j \ast$ to both sides of equation~\eqref{eq:diff-sep}. Using equation~\eqref{eq:dual1}
we obtain the equation
\begin{equation} \label{eq:d2}
\rho \, j\ast \u^{\prime\prime} = t \ast L[\u^\prime] + j\ast \f(t,\x)
\end{equation}
The Fubini theorem implies the identity 
$$\int_0^t \left[ \int_0^s f(\xi)\, \dd \xi\right] \, \dd s = \int_0^t (t - \xi) \, f(\xi) \,
\dd \xi$$
Using this identity equation~\eqref{eq:d2} can be transformed to the following form
$$\rho \, j\ast \u^{\prime\prime} = \int_0^t L[\u(s,\x) - \u(0,\x)]\, \dd s + j\ast\f$$
Differentiating the last equation with respect to time and noting the identity
$$\partial \left[ j\ast \u^{\prime\prime}\right]/\partial t = 
\partial \int_0^t j(t - s) \, \u^{\prime\prime}(s,\x)\, \dd s/\partial t = 
j(0) \, \u^{\prime\prime} + j^\prime\ast\u^{\prime\prime} $$
we get the equation 
\begin{equation} \label{eq:diff2}
j(0+)\, \u^{\prime\prime} + j^\prime\ast \u^{\prime\prime} = 
L [\u(t,\x) - \u(0,\x)] + j(0+) \, \f(t,\x) + j^\prime\ast \f
\end{equation}

Recall Bernstein's theorem for the LICM function $j^\prime$ \cite{BernsteinFunctions}:
\begin{equation} \label{eq:Bernstein}
j^\prime(s) = \int_{[0,\infty[} \e^{-r s}\, \nu(\dd r)
\end{equation}
where $\nu$ is a positive Radon measure on $[0,\infty[$ satisfying the
inequality 
\begin{equation} \label{eq:in1}
\int_{[0,\infty[} \frac{\nu(\dd r)}{1 + r} < \infty
\end{equation}
The convolution $j^\prime \ast \u^{\prime\prime}$ can now be expressed 
in terms of an integral over $[0,\infty[$:
\begin{equation}
\left[j^\prime \ast \u^{\prime\prime}\right](t,\x) = \int_{[0,\infty[} \mathbf{\phi}_r(t,\x) \, \nu(\dd r)
\end{equation}
where 
\begin{equation}
\mathbf{\phi}_r(t,\x) := \int_0^t \e^{-r (t-s)}\, \u^{\prime\prime}(s,\x)\, \dd s
\end{equation}
Note that $\mathbf{\phi}_r$ can be calculated simultaneously with $\u$ by integrating
the differential equations
\begin{equation}
\mathbf{\phi}_r^{\;\prime} + r \,\mathbf{\phi}_r - \u^{\prime\prime} = 0 \qquad r \geq 0
\end{equation}
with the initial condition $\mathbf{\phi}_r(0,\x) = 0$.

We thus end up with the equation
\begin{equation} \label{eq:semifinal}
\rho\, j(0+) \, \u^{\prime\prime} + \int_{[0\infty[} \mathbf{\phi}_r(t,\x) \, \nu(\dd r)
= L[\u(t,\x) - \u(0,\x)] + j(0+) \, \f(t,\x) + j^\prime\ast \f
\end{equation}

The integral on the left-hand side is convergent. Indeed 
$$r\, \| \mathbf{\phi}_r(t,\x)\| \leq \int_0^t r\, \e^{-r (t-s)} \, \| \,
\u^{\prime\prime}(s,\x)\| \, \dd s$$
The integrand on the right tends to 0 as $r \rightarrow \infty$ for $0 \leq s < t$, hence
almost everywhere on $[0,t]$. If $\sup_{0 \leq s \leq t} \| \u^{\prime\prime}(s,\x)\| < C 
< \infty$ then the integrand is bounded and integrable on $[0,t]$. By the Lebesgue Dominated
Convergence Theorem $\lim_{r\rightarrow \infty} r \, \mathbf{\phi}_r(t,\x) = 0$
and in view of \eqref{eq:in1} the integral in equation~\eqref{eq:semifinal} is convergent.

Suppose that the second term on the left-hand side of equation~\eqref{eq:semifinal} can 
be approximated
by a quadrature with nodes $r_l$ and weights $w_l$, $l = 1,\ldots, n$, independent of $t, \x$:
\begin{equation}
\int_{[0\infty[} \mathbf{\phi}_r(t,\x) \, \nu(\dd r) \sim \sum_{l=1}^n w_l \,
\mathbf{\phi}_{r_l}(t,\x) 
\end{equation}
and let $\mathbf{\psi}_l(t,\x) := \mathbf{\phi}_{r_l}(t,\x)$ for $l = 1,\ldots, n$.
The original IVP can be approximated by the equations
\begin{gather}
\rho\, j(0+) + \sum_{l=1}^n w_l \, \mathbf{\psi}_l = L[\u] + j(0+) \, \f(t,\x) + 
j^\prime\ast \f \qquad t \geq 0, \quad \x \in \U  \label{eq:1}\\
\mathbf{\psi}_l^{\;\prime} + r_l \, \mathbf{\psi}_l - \u^{\prime\prime} = 0 \qquad l = 
1,\ldots, n; \; t \geq 0;\; \x \in \U \label{eq:2}\\
\u(0,\x) = 0; \; \u^\prime(0,\x) = 0;\; \mathbf{\psi}_l(0,\x) = 0 \qquad \x \in \U\\
\u(t,\x) = 0 \qquad \x \in [0,\infty[\, \times\Gamma_1 \\
\n\cdot \C(\x)[\nabla\, \u(t,\x)] = 0 \qquad \x \in [0,\infty[\, \times\Gamma_2
\end{gather}

Discretize the time according to the formula $t = m\, h$, where $h > 0$ is a step size, 
and let $\u_m(\x) := \u(m\, h, \x)$, $\u^{\prime\prime}_m(\x) := \u^{\prime\prime}(m\, h,\x)$,
$\psi_{l,m}(\x) := \psi_l(m\,h,\x)$. Suppose that 
$\u_k, \psi_{l,k}$ has been calculated for $k \leq m$, $1 \leq l \leq n$.
According to a tentative numerical scheme 
equation~\eqref{eq:1} can be used do calculate $\u^{\prime\prime}_m$. Equations~\eqref{eq:2}
can be used to calculate $\psi_{l,m+1}$, while $\u_m, \u_{m-1}$ and $\u^{\prime\prime}_m$ 
can be used to calculate $\u_{m+1}$.  

The Radon measure $\nu$ can be expressed in terms of the retardation measure \cite{HanDuality}.
Apply the general integral representation of a Bernstein function
to $j(t)$ \cite{BernsteinFunctions}:
\begin{equation} \label{eq:ret}
j(s) = j(0+) + s\, N + \int_{]0,\infty[} \left[ 1 - \e^{-r s}\right] \, \mu(\dd r)
\end{equation}
where the retardation measure $\mu$ is a positive Radon measure on $]0,\infty[$ satisfying
the inequality
\begin{equation} \label{eq:in2}
\int_{]0,\infty[} \frac{r\, \mu(\dd r)}{1 + r} < \infty
\end{equation}
Equation~\eqref{eq:ret} implies equation~\eqref{eq:Bernstein} with $\nu(\dd r) =
r \, \mu(\dd r)$ for $r > 0$ and $\nu\{0\} = N$. Equation~\eqref{eq:in2} implies 
equation~\eqref{eq:in1}.

\section{Application to the Strick-Mainardi creep.}

The method presented in Section~\ref{retbased} is particularly effective if the retardation spectrum, i. e. the support of the
measures $\mu$ and $\nu$ is a finite interval. This is in particular the case for the
Strick-Mainardi creep compliance $j(t)$, defined by its Laplace transform
\begin{equation} \label{eq:SM}
\tilde{j}(p) = J_0 + M_0 \, \left[ ( 1 + \Omega/p)^\alpha - 1 \right]/\alpha
\end{equation}
where $J_0, M_0$ and $\Omega$ are some positive constants and $0 < \alpha < 1$.
The Radon measure $\nu$ is given by the formula $\nu(\dd r) = W(r)\, \dd r$, where
$W(r) := \Im \left[ \widetilde{J^\prime}\left(r \,\e^{-\ii \uppi}\right)\right]/\uppi$
which in our particular case is given by the formula is 
\begin{equation}
W(r) = \begin{cases} M_0 \frac{\sin(\alpha \, \uppi)}{\alpha \uppi}
 \left(\Omega/p - 1\right)^\alpha & r \leq \Omega \\
0 & r > \Omega
\end{cases}
\end{equation}
for $r \geq 0$. Hence
$$ j^\prime \ast \u^{\prime\prime} = \int_0^\Omega W(r)\, \mathbf{\phi}_r(t,\x)\, \dd r$$
The retardation spectrum is the interval $[0,\Omega]$. The convolution 
\begin{equation}
j^\prime\ast \u^{\prime\prime}(t,\x) = M_0\, \Omega \, \int_{-1}^1 (1 - \xi)^\alpha \,
(1 + \xi)^{-\alpha} \mathbf{\phi}_{\Omega\, (1+\xi)/2}(t,\x)\, \dd \xi
\end{equation}
can be approximated by Gauss-Jacobi quadrature \cite{StroudSecrest}:
\begin{equation}
j^\prime\ast \u^{\prime\prime}(t,\x) = M_0 \, \Omega\, \frac{\sin(\alpha \uppi)}{\alpha \uppi}
\sum_{l=1}^n w_l\, \mathbf{\phi}_{\Omega (1 + \xi_l)/2}(t,\x)
\end{equation}
where the nodes $\xi_l$, $l = 1,\ldots, n$ are the roots of a Jacobi polynomial of
degree $n$ and $w_l$, $l = 1,\ldots, n$, are the corresponding weights 
\cite{StrickMainardi,Hanyga}. The nodes $\xi_l$ and the weights $w_l$ can be
calculated by the programs presented in Sec.~2.5 of \cite{StroudSecrest}.    

A numerical approximation of the operator $L[\u]$ can be constructed by the FEM 
\cite{StrangFix} or the pseudospectral method \cite{Fornberg}. 

\section{Concluding remarks.}

Numerial integration of equations involving memory effects normally requires multiplications 
and summations 
of an ever increasing sequence of past values of each physical variable. The proposed method is based on the idea
of auxiliary variables which contain the information relevant for the future evolution
of the system. There is a continuum of such variables and their contributions should 
be summed with appropriate weights. The resulting integral can however be much easier to 
calculate than the time convolution of memory dependent variables. 

In many cases the retardation spectrum is infinite. In the case of an infinite retardation 
spectrum the integration domain can be mapped to a finite interval by a Cayley transformation.
the quadrature can also be reduced to a few nodes by applying an asymptotic estimate of
the integrand \cite{HanygaLu}. 

A Caputo fractional derivative of order $\alpha \in ]0,1]$ is given by the formula
$\D^\alpha f = \int_0^t \frac{s^{-\alpha}}{\Gamma(1-\alpha)} f^\prime(t-s)\, \dd s$
\cite{Podlubny}
and $s^{-\alpha}/\Gamma(1-\alpha)$ is a LICM function. It is therefore possible to apply 
similar methods to equations with fractional derivatives. 

In particular a viscoelastic medium can be defined in terms 
of Caputo fractional derivatives. This approach to hereditary viscoelasticity has a long history
associated with the names of Gemant, Scott-Blair, Rabotnov, Caputo and Koeller.
A detailed account of this class of viscoelastic models
can be found in \cite{Mainardi}. We shall now relate these models to the models based
on completely monotonic relaxation moduli and Bernstein creep compliances 
\cite{Bland,Molinari,HanDuality}. 
If the creep compliance is given by
the formula 
$j(s) = M_0\, s^{1-\alpha}/\Gamma(2-\alpha)$ then the convolution $j^\prime \ast \u^{\prime\prime} =
M_0\, \D^{\alpha + 1} \u$ is a Caputo derivative. Equation~\eqref{eq:diff2}
assumes the form
$$
j(0+)\, \u^{\prime\prime} + M_0\, \D^{\alpha+1}\, \u = 
L [\u(t,\x) - \u(0,\x)] + j(0+) \, \f(t,\x) + j^\prime\ast \f
$$
In this case 
$j^\prime(s) = s^{-\alpha}/\Gamma(1-\alpha) = \int_0^\infty \e^{-r t}\, 
r^{\alpha - 1}/\Gamma(\alpha) \dd r$, hence the retardation spectrum is infinite
and $W(r) = r^{\alpha-1}/\Gamma(\alpha)$. Equation~\eqref{eq:duality} in the
Laplace domain assumes the form $\tilde{j}(p) \, \tilde{g}(p) = p^{-2}$ with 
$\tilde{j}(p) = p^{\alpha-2}$, hence $\tilde{g}(p) = p^{-\alpha}$ and
$g(s) = s^{\alpha-1}/\Gamma(\alpha)$. Consequently the constitutive equation assumes 
the form $\upsigma(t,\x) = \D^{1-\alpha} \C(\x)[\nabla \u]$.

\end{document}